# Two-dimensional Halide Perovskites: Tuning Electronic Activities of Defects

Yuanyue Liu,[*,1,2] Hai Xiao[1] and William A. Goddard III[*,1]

[1]*Materials and Process Simulation Center*
[2]*the Resnick Sustainability Institute*
*California Institute of Technology, Pasadena, CA 91125, USA*

[*]Correspondence to: yuanyue.liu.microman@gmail.com and wag@wag.caltech.edu

**Abstract:** Two-dimensional (2D) halide perovskites are emerging as promising candidates for nano-electronics and optoelectronics. To realize their full potential, it is important to understand the role of those defects that can strongly impact material properties. In contrast to other popular 2D semiconductors (e.g. transition metal dichalcogenides $MX_2$) for which defects typically induce harmful traps, we show that the electronic activities of defects in 2D perovskites are significantly tunable. For example, even with a fixed lattice orientation, one can change the synthesis conditions to convert a line defect (edge or grain boundary) from electron acceptor to inactive site without deep gap states. We show that this difference originates from the enhanced ionic bonding in these perovskites compared with $MX_2$. The donors tend to have high formation energies, and the harmful defects are difficult to form at a low halide chemical potential. Thus we unveil unique properties of defects in 2D perovskites and suggest practical routes to improve them.

**Main text**: Halide perovskites have attracted great interest due to their low cost and high efficiency for solar cell applications[1]. Recently two-dimensional (2D) halide perovskites have been realized experimentally and demonstrated to have attractive properties. These materials have thicknesses of just one to few unit-cell(s), with an $A_2BX_4$ stoichiometry (where X = Halides, B = group-14 elements, and A = long-chain organic molecules such as $C_4H_9NH_3$) in contrast to $ABX_3$ for 3D perovskites [2-5]. The excellent properties of 2D perovskites combined with their ease of fabrication render them promising for nano-device applications. For example, they exhibit strong light absorption and photoluminescence at room temperature [2, 5], making them interesting for photovoltaics and light emitters [6-10]. In addition, the high mobility of charge carriers[11-15] in thin film perovskites renders them promising candidates for solution-processed field-effect transistors[12, 13, 15].

To optimize the 2D perovskites, it is important to understand the impact of defects on the material properties and device performance. Although defects in 3D perovskites[16-19] and in other 2D materials (graphene[20, 21], boron nitride[22, 23], transition metal dichalcogenides[24-26], black phosphorous[27]) have been studied extensively, little is known about defects in the emerging 2D perovskites. Here we report first-principles studies to answer such questions as:

- what are the electronic properties of defects in 2D perovskites?

- how are they different from other 2D semiconductors (especially transition metal dichalcogenides, which are also hetero-elemental semiconductors) and 3D perovskites?
- how can we control defects to optimize the device performance?

We performed Density functional theory (DFT) using the Vienna Ab-initio Simulation Package (VASP)[28, 29] with projector augmented wave (PAW) pseudopotentials[30, 31]. We employed the Perdew-Burke-Ernzerhof (PBE) exchange-correlation functional[32] in most systems. For comparison, we also calculated the band gap using the HSE functional[33] with spin-orbit coupling (SOC). The plane-wave cut-off energy is 400 eV, and the systems are fully relaxed until the final force on each atom becomes less than 0.01 eV/Å. In order to reduce computational costs, we use Rb to represent the long-chain organic molecules (A). This is based on the considerations that the main role of A in the electronic structures of 3D perovskites is to donate one electron into the host[34]. Although Rb has a smaller size than A and hence leads to a different lattice parameter, it does not affect our main conclusions about the defect properties, as explained below.

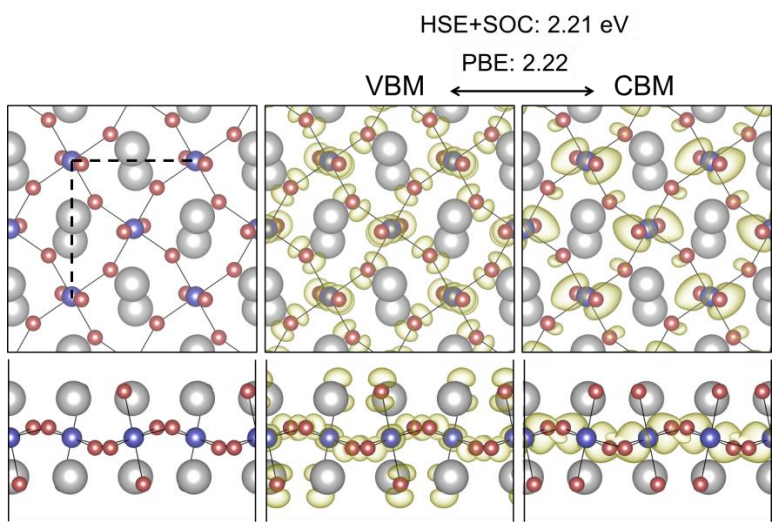

**Figure 1**. Atomic structure (left) of 2D perovskite and charge density distributions (middle and right) of the band edge states, shown in both top (upper panels) and side (lower panels) views; blue: Pb; reb: I; grey: Rb. The band gap is calculated to be ~ 2.2 eV with both PBE and HSE+SOC flavors of DFT.

Fig. 1 shows the atomic structure of 2D $Rb_2PbI_4$. The octahedra are tilted, along both in-plane and out-of-plane directions. Using the PBE functional without SOC, we calculate a band gap of 2.22 eV, which is consistent with the band gap of 2.21 eV that we obtain from the more accurate HSE + SOC method. This suggests that PBE is acceptable for studying defect properties, as previously noted for 3D perovskites[16-19]. The spatial distributions of the band edge states show that the valence band maximum (VBM) is mainly composed of Pb and I states, while the conduction band minimum (CBM) is dominated by Pb states, with Rb not contributing to the band edges. This absence of Rb components near the band edges further validates our choice of

Rb to mimic A for studying the defect electronic properties. These features are similar to those of 3D perovskites[34], indicating a similar electronic origin despite the apparently different stoichiometry. On the other hand, these band edge compositions are very different from 2D $MX_2$, whose VBM and CBM are both dominated by M d states split in a ligand field[35]. The spatial separation of VBM and CBM onto anions and cations suggests that the 2D perovskites possess more ionic bonding than $MX_2$.

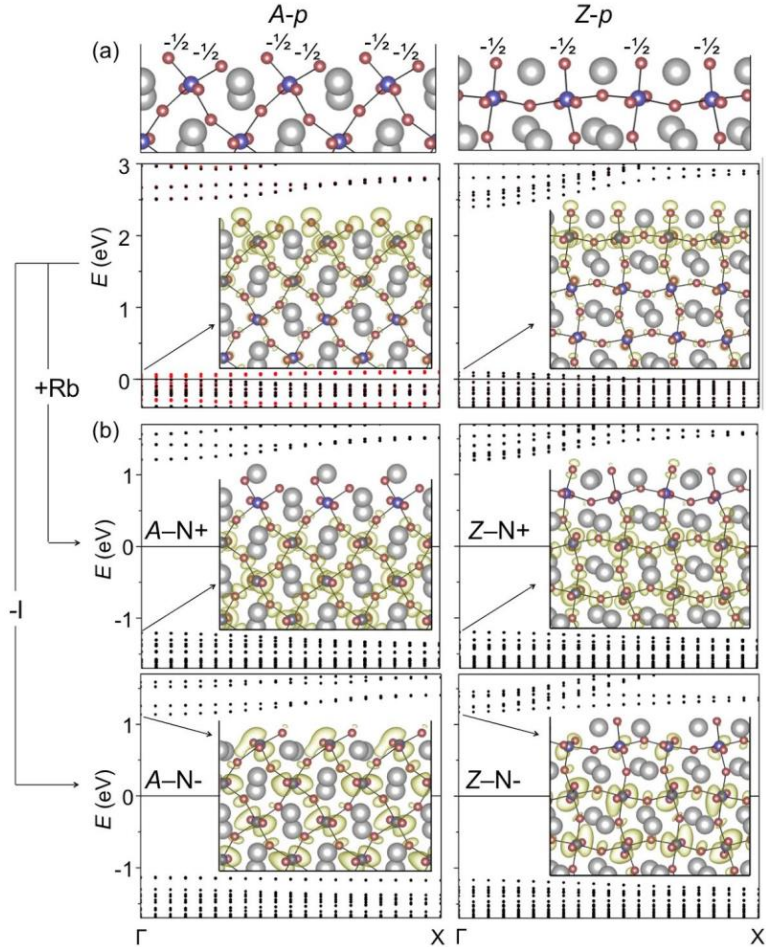

**Figure 2**. Edges in 2D perovskite and their electronic structures. '*A*' indicates 'armchair' orientation and '*Z*' indicates 'zigzag'. The suffix denotes the specific structure: '-*p*' indicates that the edge creates acceptor levels, and '-N' means that the edge is inactive ('neutral'). Spin-polarized states are shown in different colors in the band structures, and charge density distributions of the states indicated by arrows are shown in the inset. (a) shows the cases of *A–p* and *Z–p* edges, and (b) shows the rest. See SI for more edge structures.

Defects in 2D $MX_2$ (point defects, edges, grain boundaries) typically create deep electronic levels inside the band gap[24-26, 36], which could trap/scatter/recombine charge carriers making them generally harmful for many (opto)electronic applications. These deep levels are difficult to eliminate by local structural variations without introducing additional chemical species[26, 37-39], due to the difficulty in restoring the original ligand field. However, for 2D perovskites, it is possible to recover the charge transfer characteristics of the ionic bonding by manipulating the ratio of cations and anions at the defect sites, thereby tuning their electronic levels.

Indeed, our study of the edges – an important type of line defects – validates this speculation. Figure 2 shows two representative edge orientations: armchair (*A*) and zigzag (*Z*) directions. The *A* edge orientation is along the axis of the primitive cell, and the *Z* is along the diagonal direction. Each edge orientation can have various structures, denoted by the suffix (e.g. –p, –N+, –N-). The *A–p* edge, which has the same coordination of Pb and Rb as in the lattice (i.e. four I atoms close to Pb, with Rb atoms up and down in the centers of the polygons), creates shallow acceptor levels located along the edge, as shown by the band structure and the charge density distribution in Fig. 2a. These edge states can be partially occupied by thermally ionized electrons from the lattice valance band, generating free holes in the lattice (hence denoted as *A–p*).

The acceptor states originate from the non-fully filled valence bands created by the surplus I atoms at the *A–p* type edge. This can be understood by counting the charges for the local stoichiometry. The I are distributed in three layers (Fig. 1). In the top and bottom layers of the ideal lattice, each I receives ¼ electron per neighboring Rb from four Rb neighbors, thus the charges are balanced. This is different from the middle layer, where each I receives 1/2 electron per neighboring Pb from two Pb neighbors, neutralizing the middle layer. However, At the *A–p* edge, although the top and bottom layers are charge balanced, the outmost I atoms in the middle layer lack ½ charge per I due to the missing Pb (Fig 2a), which gives rise to the acceptor states. Although there are other ways to count the charges, they all should lead to the same conclusion.

The above analysis suggests that adding one Rb atom to the *A–p* edge might saturate the two outmost I. Indeed, our calculations of band structure and charge density distribution (Fig. 2b, A–N+, where 'N' denotes 'neutral', and + means adding atoms to the previous *A–p* edge) show that the acceptor levels disappear from the band gap, leading to the absence of edge states. Therefore, this edge is relatively inactive with regard to the lattice electronic properties. Alternatively, removing the unsaturated I atoms, also results in an electronically inactive edge A–N- (Fig. 2b; - means removing atoms from the previous *A–p* edge). We can construct edges with even more cations or fewer anions (see the SI for structures), that would create donor levels (hence denoted as *A–n*) to generate free electrons in the lattice conduction band. However, as shown below, we find that these edges are very unstable (very high formation energies).

Similarly, the *Z* edge provides opportunities, to stabilize either electron acceptor (Fig. 2a, *Z–p*) or inactive (Fig. 2b, *Z*–N+, *Z*–N-) states, depending on the stoichiometry at the edge. It is also unlikely to be donor due to the high formation energies of the Z edge structures that could create donor levels. These edge properties are very different from those of $MX_2$, which always exhibit deep levels independent of structural variations[36], demonstrating the unique electronic structure of 2D perovskites. These observations suggest that, even for fixed edge directions, the electronic activity can still be tuned by varying synthesis conditions.

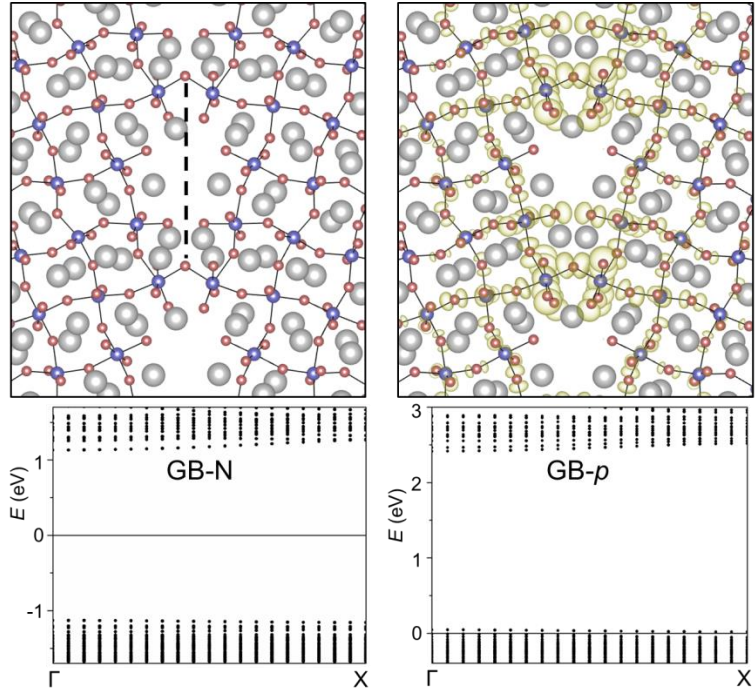

**Figure 3**. Gain boundaries in 2D perovskite and their electronic structures. The dashed line in the left panel shows the periodic length. The charge density distributions of the acceptor levels are shown in the right panel. See SI for more grain boundary structures.

Grain boundaries provide another common type of line defects, usually formed when the edges of two mis-orientated grains join together during growth. Figure 3 shows an example grain boundary constructed by connecting two edges with a shared I atom. We choose the kinked edges that contain both *A* and *Z* segments to represent a general case. By varying the number of Rb atoms, shallow acceptor levels can be created or eliminated. It is energetically unfavorable to have surplus cations (as shown below), so we expect the grain boundary is unlikely to provide electron donor states. These grain boundary properties are very different from those of $MX_2$, which always render deep levels regardless of structural variations[24].

We find a similar charge-balance-controlled electronic activity for point defects in 2D perovskites, as shown in Fig. 4. Although a Rb vacancy ($V_{Rb}$) creates an acceptor level, a neighboring $V_I$ (hence converting it to $V_{RbI}$) could eliminate this gap state. Similarly, $V_{PbI2}$ does not exhibit deep levels. Such defects have also been found to be electronically inactive in 3D perovskites[19]. Most of point defects have the electronic behavior expected for a typical ionic semiconductor. For example, cation vacancies/anion interstitials usually generate acceptor levels, while anion vacancies/cation interstitials typically create donor states. These point defect properties are very different from those of $MX_2$, In the latter case, a cation vacancy ($V_M$) generates deep acceptor levels, while the stoichiometric vacancies ($V_{MX2}$) produce more gap states[25].

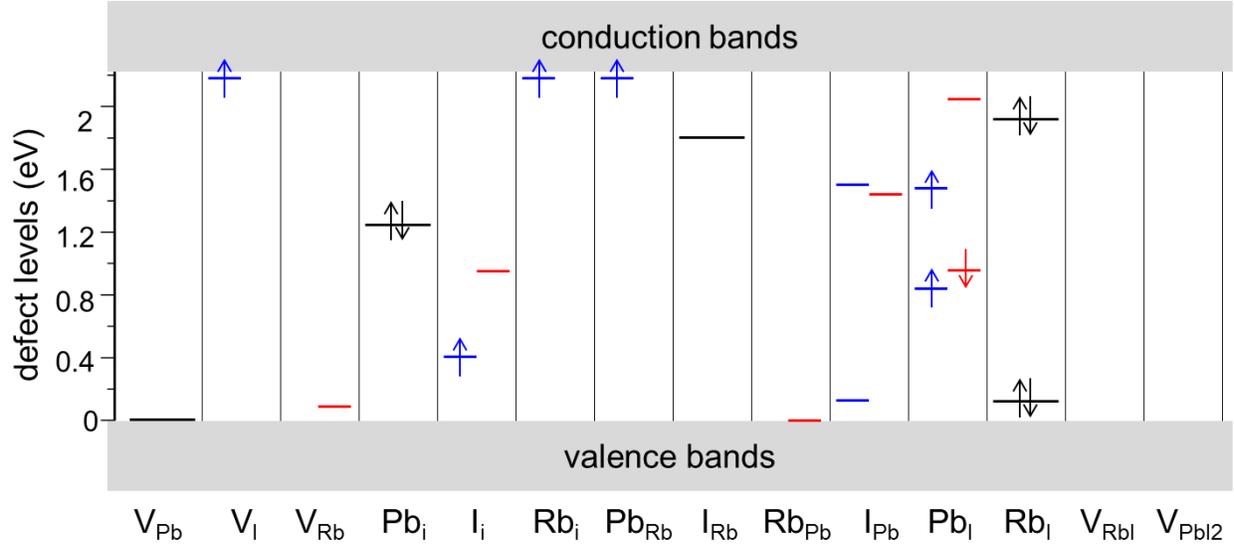

**Figure 4**. Electronic levels of point defects in 2D perovskite. A long bar denotes two degenerate states, while a short bar stands for a single state. Spin polarized states are shown in different colors, and the occupied states are marked by arrows.

In order to identify optimal conditions for growth of materials that suppress harmful defects, we examine the formation energies ($E_f$) following the method used for 3D perovskites[16]. Thermodynamic equilibrium condition requires:

$2\mu_{Rb} + \mu_{Pb} + 4\mu_I = \mu_{Rb2PbI4}$ (1)

where μ is the chemical potential. To avoid phase separation, the following constraints must be satisfied:

$\mu_{Rb} < \mu_{Rb\text{-bulk}}$ (2)

$\mu_{Pb} < \mu_{Pb\text{-bulk}}$ (3)

$\mu_I < \mu_{I2}/2$ (4)

$\mu_{Rb} + \mu_I < \mu_{RbI}$ (5)

$\mu_{Pb} + 2\mu_I < \mu_{PbI2}$ (6)

Substituting (1) into (2) and (5), we get:

$\mu_{Pb} + 4\mu_I > \mu_{Rb2PbI4} - 2\mu_{Rb\text{-bulk}}$ (7)

$\mu_{Pb} + 2\mu_I > \mu_{Rb2PbI4} - 2\mu_{RbI}$ (8)

where $\mu_{Rb2PbI4}$, $\mu_{PbI2}$, $\mu_{RbI}$, $\mu_{Rb\text{-bulk}}$, $\mu_{Pb\text{-bulk}}$ and $\mu_{I2}$ can be approximated by the internal energies of the corresponding condensed phases. We find that (7) or (2) is automatically satisfied when (3),

(4), (6) and (8) are met. Hence (3), (4), (6) and (8) together define a range of ($\mu_{Pb}$, $\mu_I$) where the 2D perovskite is thermodynamically stable. Since the $E_f$ depends linearly on $\mu$, the maximum and minimum of $E_f$ should fall on the corners of the phase boundaries. Therefore Fig. 5 shows $E_f$ along the two boundary lines: $\mu_{Pb} + 2\mu_I = \mu_{PbI2}$, and $\mu_{Pb} + 2\mu_I = \mu_{Rb2PbI4} - 2\mu_{RbI}$ (or $\mu_{Rb} + \mu_I = \mu_{RbI}$).

The thermodynamic equilibrium concentration of defects ($n$) in the materials can be estimated by:

$n \sim e^{(-E_f/k_BT)} /S$      (9)

where $S$ is the area of the primitive cell, and $T$ is temperature. The experimentally grown 2D perovskites typically exhibit sizes less than 10 μm, and T is usually below 100 °C[2]. Based on (9), we estimate that defects with $E_f < 0.62$ eV would likely form under these experimental growth conditions. Therefore we use this value as a criterion to judge if $E_f$ is 'high' or 'low'. Although $V_{RbI}$ and $V_{PbI2}$ generally have a low $E_f$ (Fig. 5a), they are electronically inactive and hence have limited impact on the lattice properties. The dominating defects at high $\mu_I$ are those with surplus anions or deficient cations, such as $I_{Rb}$, $I_i$, and $I_{Pb}$ (Fig. 5a). These defects create deep levels (Fig. 4) and hence are harmful to many applications. Fortunately, their $E_f$ increase to a high level as $\mu_I$ decreases. On the other hand, the $E_f$ for defects with surplus cations/deficient anions still remains high at low $\mu_I$. Therefore using synthesis conditions that lower $\mu_I$, should reduce the total concentration of harmful defects.

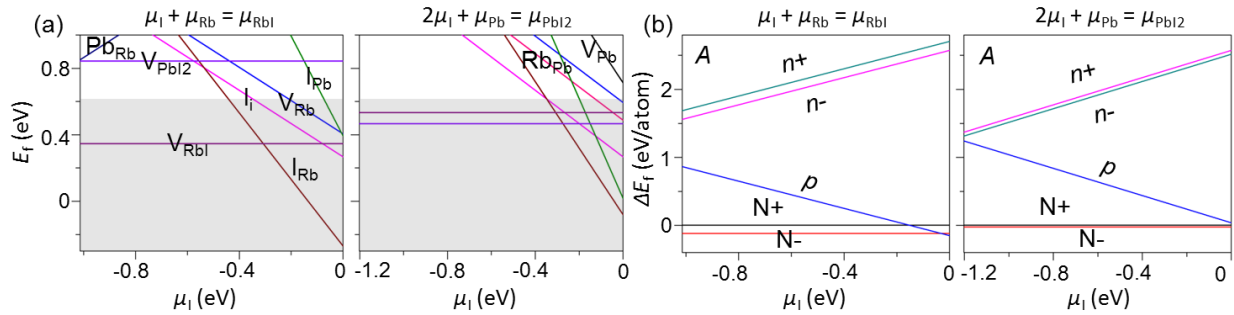

**Figure 5**. Formation energies of point defects (a), and line defects (b) in 2D perovskite, as a function of I chemical potential (with respect to that of $I_2$ molecule) along phase boundaries (see the text). For line defects, edges along $A$ orientation are shown here as an example, and the rest can be found in the SI; the energies are referred to that of $A$–N+. Shadowed regions mark the point defects that would likely form in a 10μm size square sheet grown at 100 °C in thermodynamic equilibrium.

We find that the line defects exhibit a similar behavior for $E_f$. Fig. 5b shows the $E_f$ for various edge structures along $A$ orientation, with respect to that of $A$–N+ (the $E_f$ for edges along $Z$ orientation and grain boundaries can be found in the SI). At high $\mu_I$, the edge that creates acceptor levels, with surplus I ($A$–$p$), possesses an $E_f$ comparable with those of inactive edges ($A$–N+ and $A$–N-). However, it becomes unfavorable at low $\mu_I$. In contrast, the edges that create donor levels with surplus cations/deficient anions ($A$–$n$+ and $A$–$n$-, see structures in the SI), exhibit a high $E_f$ in the whole range of $\mu_I$, and therefore are unlikely to form (as mentioned above).

To check whether the trends of $E_f$ can be generalized to other 2D perovskites with different chemical compositions, we calculate the $E_f$ for point defects in $(CH_3NH_3)_2SnBr_4$ as a test example. As shown in Fig. S6, we find again that a low $\mu_{Br}$ can decrease the total concentration of harmful defects, therefore confirming the generality of the trends.

The behavior of $E_f$ in 2D perovskites is different from that in 3D perovskites, where point defects with surplus cations/deficient anions can have a low $E_f$ at low $\mu_I$, rendering n-doping of the host [16, 17]. This n-doping is unlikely to exist in 2D case, because of the high $E_f$ for donors across the whole range of chemical potential. Note that the same calculation methods were used to study the defects in 3D case, i.e. PBE functional with plane-wave basis sets, allowing for direct comparison. For grain boundaries in 3D perovskites, theoretical analyses suggested that they do not create deep levels, due to the strong coupling between Pb s orbitals and I p orbitals and the large atomic size of Pb[17, 18]. We show here that these previous results arose because the grain boundary models chosen were all neutral (charge balanced), making them electronically inactive as explained above for 2D cases. Considering that both donor- and acceptor-like point defects can form in 3D perovskites, we anticipate that the grain boundaries can also have surplus/deficient cations/anions, making them donors/acceptors depending on the $\mu_I$. This is different from the grain boundaries in our 2D case, which are unlikely to be donors. In addition, theoretical analyses suggested that deep-level defects are difficult to from in 3D perovskites and the dominating defects all have shallow states[16, 17]; this contrasts with defects in 2D perovskites, where deep-level defects (e.g. $I_i$, $I_{Rb}$) can form easily at high-$\mu_I$ conditions.

This study demonstrates that defects in ionic semiconductors can be tuned to be less harmful in general, providing a guideline to design new 2D semiconductors. It also explains the experimental observation of relatively high quantum efficiency in 2D perovskites, and suggests ways to further improve it. A common way to adjust the chemical potential is to change the concentration of reactants. For example, recent experiments use $PbX_2$ and Cs-oleate to synthesize $CsPbX_3$ nanostructures, creating a Pb-rich (or I-poor) environment[40, 41]. Besides the intrinsic defects which are the focus of this work, extrinsic defects would also play an important role in the electronic properties. A major source is the solvent residues adsorbed on the surface. The ionic contaminants could induce *n*- or *p*-doping, while neutral adsorbates should have less impact. Given that 3D perovskite is not very stable in the ambient conditions, one would expect a similar issue for 2D perovskite. Particularly, humidity could have a strong effect on the material. This could be mitigated by using encapsulation techniques (e.g. using h-BN to seal the material/device[42]), or choosing hydrophobic organic cations[43].

In summary, we use first-principles calculations to predict unique properties of defects in 2D perovskites. The line defects with fixed orientation can be tuned from electron acceptors to inactive sites by varying synthesis conditions, while donors are energetically unfavorable. This is consistent with the trends of point defects formation. The optimal synthesis conditions are also identified.

**Supporting Information:**

Computational details, more line defect structures, energies of $Z$ edges and grain boundaries, energies of point defects in 2D $MA_2SnBr_4$.

**Acknowledgements:**

YL thanks discussions with Prof. Wan-Jian Yin, and acknowledges the support from Resnick Prize Postdoctoral Fellowship at Caltech. HX and WAG were supported by the Joint Center for Artificial Photosynthesis, a DOE Energy Innovation Hub, supported through the Office of Science of the U.S. DOE under Award No. DE-SC0004993. This research was also supported by NSF (CBET-1512759, program manager: Robert McCabe), DOE (DE FOA 0001276, program manager: James Davenport). This work used computational resources of National Energy Research Scientific Computing Center, a DOE Office of Science User Facility supported by the Office of Science of the US DOE under Contract DE-AC02-05CH11231, and the Extreme Science and Engineering Discovery Environment (XSEDE), which is supported by NSF grant number ACI-1053575.

TOC:

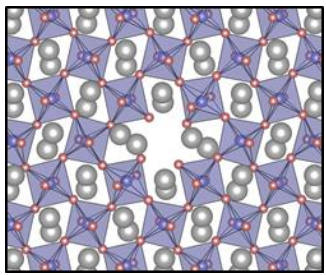